\documentclass[]{pasj01}
\draft

\Received{}
\Accepted{}
 
\begin{document} 

\title{{\it TESS} Photometry of the Eclipsing $\delta$ Scuti Star AI Hydrae }

\author{Jae Woo \textsc{Lee}\altaffilmark{1}%
}
\altaffiltext{1}{Korea Astronomy and Space Science Institute, Daejeon 34055, Republic of Korea}
\altaffiltext{2}{Institute for Astrophysics, Chungbuk National University, Cheongju 28644, Republic of Korea}
\altaffiltext{3}{DTU Space, National Space Institute, Technical University of Denmark, Elektrovej 327, DK-2800 Lyngby, Denmark}
\altaffiltext{4}{Brorfelde Observatory, Observator Gyldenkernes Vej 7, DK-4340 T\o{}ll\o{}se, Denmark}
\email{jwlee@kasi.re.kr}

\author{Kyeongsoo \textsc{Hong}\altaffilmark{2}}
\author{Martti H. \textsc{Kristiansen}\altaffilmark{3,4}}

{}

\maketitle

\begin{abstract}
AI Hya has been known as an eclipsing binary with a monoperiodic $\delta$ Sct pulsator. We present the results from its {\it TESS} 
photometry observed during Sector 7. Including our five minimum epochs, the eclipse timing diagram displays the apsidal motion with 
a rate of $\dot{\omega}$ = 0.075$\pm$0.031 deg year$^{-1}$, which corresponds to an apsidal period of U = 4800$\pm$2000 years. 
The binary star model represents that the smaller, less massive primary component is 427 K hotter than the pulsating secondary, and 
our distance of 612$\pm$36 pc is in good agreement with the $Gaia$ distance of 644$\pm$26 pc. We subtracted the binary effects from 
the observed {\it TESS} data and applied a multifrequency analysis to these residuals. The result reveals that AI Hya is multiperiodic 
in its pulsation. Of 14 signals detected, four ($f_1$, $f_2$, $f_3$, $f_6$) may be considered independent pulsation frequencies. 
The period ratios of $P_{\rm pul}/P_{\rm orb}$ = 0.012$-$0.021 and the pulsation constants of $Q$ = 0.30$-$0.52 days correspond to 
$\delta$ Sct pulsations in binaries. We found that the secondary component of AI Hya pulsates in both radial fundamental $F$ modes 
($f_2$ and $f_3$) and non-radial $g_1$ modes with a low degree of $\ell$ = 2 ($f_1$ and $f_6$). 
\end{abstract}

\section{Introduction}

Eclipsing binaries (EBs) are a primary source of empirical measurements of stellar properties. In particular, detached 
double-lined EBs provide accurate determinations of absolute masses, radii, and luminosities from the simultaneous analysis 
of radial velocities (RVs) and light curves (Hilditch 2001; Torres et al. 2010). The fundamental data are used to test 
stellar evolution models and to determine geometric distance to the binary systems. EBs with pulsating components are extremely 
valuable, because more reliable modeling of stellar pulsations allows the exploration of internal structure and rotation. 
Asteroseismic studies are dependent on external determinations of surface gravity and effective temperature, which can be 
independently measured from binary modeling. When pulsating stars are present in EBs, the synergy between them greatly 
improves our understanding about stars. 

To develop this subject, we have been looking for pulsating stars in EBs and studying their physical properties, utilizing 
{\it Kepler} satellite photometry and ground-based spectroscopy (e.g., Lee \& Park 2018; Lee et al. 2020). Recently, 
the 2-min cadence observations of the Transiting Exoplanet Survey Satellite ({\it TESS}; Ricker et al. 2015) have proved very 
helpful for asteroseismology of the pulsating EBs showing multiperiodic oscillation with low amplitudes (Lee et al. 2019; 
Antoci et al. 2019). This work focuses on AI Hya (TIC 455178154, BD+00 2259, TYC 196-626-1; $T_{\rm p}$ = $+$9.035; 
$V\rm_T$ = $+$9.40, $(B-V)\rm_T$ = $+$0.45), which is known to be an eccentric detached EB with a binary orbital period 
of about 8.29 days (Lause 1938; Busch 1970). 

From the $uvby$ light curves, J\o rgensen \& Gr\o nbech (1978) obtained the photometric elements of AI Hya, which indicate 
this system to be an eccentric EB with an orbital eccentricity of $e$ = 0.230 and an argument of periastron of $\omega$ = 245.6 deg. 
Further, they reported that the program target is a pulsating EB with a period of 0.13803 days and a semi-amplitude of 0.01 mag and 
that the pulsation corresponds to a first overtone radial mode of $\delta$ Sct stars. The double-lined RVs of the binary star were 
measured by Popper (1988). He derived the velocity semi-amplitudes of $K_1$ = 90.1 km s$^{-1}$ and $K_2$ = 83.1 km s$^{-1}$, and 
calculated the absolute parameters of each component combined with the photometric results of J\o rgensen \& Gr\o nbech (1978). 
On the other hand, Khaliullin \& Kozyreva (1989) carried out incomplete photoelectric observations of the primary and 
secondary eclipses, and obtained an apsidal motion rate of $\dot{\omega}$ = 0.029$\pm$0.049 deg year$^{-1}$ based on the comparison 
with the light curves of J\o rgensen \& Gr\o nbech (1978). Here, we present and discuss the binary and pulsational properties of 
AI Hya using the high-precision space photometry from the {\it TESS} mission.

\section{Observations}

AI Hya was observed by the {\it TESS} mission during Sector 7 in 2-min cadence mode. The observations were collected by camera 1 
from January 8 to February 1 2019 (BJD 2,458,491.63 $-$ 2,458,516.09). There is no plan to observe the binary star in 
other sectors during the {\it TESS} primary mission\footnote{https://heasarc.gsfc.nasa.gov/cgi-bin/tess/webtess/wtv.py}. We obtained 
the simple aperture photometry (SAP) data from MAST\footnote{https://archive.stsci.edu/} and omitted seven obvious outliers using 
visual inspection and the 5$\sigma$ criterion. A total of 16,397 individual measurements were used for this work. The raw SAP data 
were detrended and normalized by fitting a second-order polynomial to the outside-eclipse light curve (Lee et al. 2017). 
We converted the normalized fluxes to magnitudes by requiring a {\it TESS} magnitude of $+$9.035 at maximum light. In Figure 1, 
the full time-series data of AI Hya is presented as magnitude {\it versus} BJD. We can see that there are short-period oscillations 
in the primary minima, as well as in outside eclipses. This variability implies that the secondary component is a pulsating star 
as suggested by J\o rgensen \& Gr\o nbech (1978).

\section{Binary Modeling}

\subsection{Eclipse Timings}

From the {\it TESS} times-series data, we obtained five mid-eclipse times and their errors with the method of 
Kwee \& van Woerden (1956) based on observations during each minimum. This method requires no assumptions about the light curve 
morphology except for symmetrical eclipses and does not depend on binary parameters. Our minimum epochs are listed in Table 1 
together with those collected from the literature, where Min I and Min II represent the primary and secondary minima, respectively. 
Because the literature data were given as HJD alone, we transformed their time stamps into TDB-based BJD by using online applets\footnote{http://astroutils.astronomy.ohio-state.edu/time/} 
(Eastman et al. 2010). The $O-C_1$ timing residuals of AI Hya were computed using the linear ephemeris of Kreiner et al. (2001): 
\begin{equation}
 C_1 = \mbox{BJD}~ 2,441,726.1433 + 8.2896735E,
\end{equation}
where the reference epoch BJD 2,441,726.1433 was converted from HJD 2,441,726.1428. The eclipse timing diagram constructed with 
ephemeris (1) is displayed in Figure 2, in which the filled and open symbols are Min I and Min II, respectively. 

As shown in this diagram, the timing residuals for the primary eclipses are approximately 180$^\circ$ out of phase with those for 
the secondary eclipses. Because our program target is an eccentric EB, the eclipse timing variation may originate from apsidal motion. 
This can be represented by the ephemeris-curve equation of Gim\'enez \& Bastero (1995), which has five independent variables 
($T_0$, $P_s$, $e$, $\dot{\omega}$, $\omega_0$). The Levenberg-Marquardt technique (Press et al. 1992) was used to evaluate 
the apsidal motion elements. At this point, the initial values of $e$ and $\omega_0$ were taken from the light curve solutions 
given in Section 3.2. As a consequence, we measured the observed rate of apsidal motion of 
$\dot{\omega}$ = 0.075$\pm$0.031 deg year$^{-1}$ and hence its long period of $U$ = 4800$\pm$2000 years. The result is plotted in 
Figure 2 and summarized in Table 2. The $O-C_2$ residuals from the best-fitting elements are given in the fifth column of Table 1. 

To phase the {\it TESS} observations of AI Hya, we introduced all primary times of minima into a linear least-squares fit 
and found the following ephemeris:
\begin{equation}
\mbox{Min I} = \mbox{BJD}~ 2,458,496.34577(16) + 8.2896499(12)E.
\end{equation}
The 1$\sigma$-errors for the coefficients of the equation are given in the parentheses. The observed light curve phased with 
ephemeris (2) are plotted as gray circles in the top panel of Figure 3 as a normalized flux {\it versus} orbital phase.

\subsection{Light Curve Synthesis} 

The {\it TESS} light curve of AI Hya presents almost flat light maxima and seems to display a total eclipse at primary minimum. 
Further, the secondary minimum is displaced to an orbital phase of about 0.447. These mean that AI Hya belongs to the class 
of eccentric detached EBs. To get the binary parameters of the program target, all {\it TESS} observations were modeled using 
version 2007 of the Wilson-Devinney binary code (Wilson \& Devinney 1971, van Hamme \& Wilson 2007; hereafter W-D). 
As the binary system is well detached, we used the W-D code in mode 2 (Wilson \& Biermann 1976). In this article, the subscripts 1 
and 2 denote the primary and secondary stars being eclipsed at Min I and Min II, respectively. 

In this synthesis, we fixed the spectroscopic mass ratio of $q$ = 1.084$\pm$0.012 measured by Popper (1988). The effective 
temperature of the more massive secondary component was initialized to be $T_{\rm eff,2}$ = 6860 K, according to the color index 
of $(b-y)$ = $+$0.24 (J\o rgensen \& Gr\o nbech 1978; Popper 1988) and the color-temperature relations in Pecaut \& Mamajek (2013). 
The logarithmic bolometric ($X$, $Y$) and monochromatic ($x$, $y$) limb-darkening coefficients were taken from the values of 
van Hamme (1993). The bolometric albedos ($A$) and the gravity-darkening exponents ($g$) were set to be standard values of $A_1$ = 1.0 
and $g_1$ = 1.0, and $A_2$ = 0.5 and $g_2$ = 0.32. In Table 3, the parentheses signify the fitted parameters and their errors. 
The differential correction program was iterated until the corrections to the free parameters were lower than their standard deviations. 

The binary modeling was carried out in a method analogous to that for the {\it TESS} target TIC 309658221 (Lee et al. 2020). 
The model parameters for the observed {\it TESS} data are given as Model 1 in columns (2)$-$(3) of Table 3. The synthetic light curve 
appears as the blue solid curve in the top panel of Figure 3, and the corresponding residuals are displayed in the middle panel of 
the figure. To reduce pulsation effects in the binary parameters, we removed the pulsation frequencies 
discussed in the next section from the observed data. The pulsation-subtracted {\it TESS} data were solved with the W-D binary code 
using the Model 1 parameters as initial values. The result is presented as Model 2 in columns (4)$-$(5) of Table 3, and displayed 
in Figure 3. This synthesis indicates that the hotter but less massive primary fills 36 \% of its inner Roche lobe, while 
the larger secondary star fills 46 \%. Here, the fill-out factor is defined as $\Omega_{\rm in}/\Omega_{1,2}$. For the error estimates 
in the adjusted parameters, we divided the {\it TESS} data into 15 subsets and separately modeled them (cf. Koo et al. 2014). Then, 
the standard deviations of each parameter were computed, and the 1$\sigma$-values were adopted as the parameter errors presented in 
Table 3.

\section{Pulsational Characteristics}

The light curve residuals of AI Hya, obtained by subtracting the binary effects from the observed {\it TESS} data, are illustrated 
in Figure 4 as magnitude {\it versus} BJD. We can clearly see the multiperiodic oscillations on a timescale of hours in 
the eclipse-subtracted data. Considering our absolute parameters presented in the following section, both components of AI Hya 
reside inside the $\delta$ Sct and $\gamma$ Dor instability strips in the Hertzsprung-Russell (HR) diagram (e.g., Lee 2016). 
J\o rgensen \& Gr\o nbech (1978) reported that the larger and more massive secondary star is pulsating at a frequency of $f_{0}$ 
= 7.2448 day$^{-1}$. The {\it TESS} time-series data exhibit total eclipses and oscillations at the primary minima. These indicate 
that the smaller primary star is completely eclipsed by the cooler secondary and the latter component is the main source of 
the light variations. 

To search for the pulsation frequencies of AI Hya, we performed a multifrequency analysis of the out-of-eclipse light residuals. 
The PERIOD04 software package (Lenz \& Breger 2005) was used to compute the amplitude spectrum up to the Nyquist frequency 
$f_{\rm Ny}$ $\simeq$ 360 day$^{-1}$. According to the successive and simultaneous prewhitening described by Lee et al. (2014), 
we calculated the signal to noise amplitude ratio (S/N) for each frequency peak and detected 14 significant frequencies by 
adopting the empirical threshold of S/N $>$ 4.0 (Breger et al. 1993). The results from this process are listed in Table 4. 
The amplitude spectra for AI Hya before and after prewhitening the first 3 frequencies, and then after all 14 frequencies, are 
displayed in the top to bottom panels in Figure 5, respectively. The synthetic curve computed from the 14-frequency fit is presented 
as a solid line in the lower panel of Figure 4. 

As shown in Figure 5 and Table 4, almost all significant signals for AI Hya lie in the frequency range of 5$-$16 day$^{-1}$, 
but we could find no credible periodicity near the $f_{0}$ frequency detected by J\o rgensen \& Gr\o nbech (1978). 
Within the frequency resolution 1.5/$\Delta T$ = 0.061 day$^{-1}$, where $\Delta T$ = 24.5 days is the observation time span 
(Loumos \& Deeming 1978), we carefully identified possible combination frequencies and orbital harmonics. The results are presented in 
the last column of Table 4. The low-frequency signal of $f_4$ = 0.1205 day$^{-1}$ appears to be the orbital frequency of $f_{\rm orb}$ 
= 0.1206 day$^{-1}$, which could be due to insufficient removal of the eclipses in the observed data. Located in the typical range of 
4$-$80 day$^{-1}$ for $\delta$ Sct stars (Breger 2000), the four frequencies of $f_1$, $f_2$, $f_3$, and $f_6$ may be 
independent pulsations originating from the more massive secondary component.

\section{Discussion and Conclusions}

In this article, we report the {\it TESS} photometry of the double-lined EB system AI Hya exhibiting total eclipses and multiperiodic 
oscillations. The binary modeling represents that the eclipse timings vary due to apsidal motion and that our target star is in 
an eccentric-orbit, detached configuration with parameters of $e$ = 0.234, $\omega$ = 250.0 deg, $i$ = 89.23 deg, 
$T_{\rm eff,1}$ = 7291 K, and $T_{\rm eff,2}$ = 6864 K. The primary and secondary components fill their inner Roche lobe by 36 \% and 
46 \%, respectively. Combining our Model 2 parameters for the {\it TESS} data and the spectroscopic elements of Popper (1988), 
we derived the absolute dimensions for AI Hya given in Table 5. The temperature error was assumed to be about 200 K, which is 
the difference between Popper's temperatures and ours. The luminosities and the bolometric magnitudes were obtained by applying 
the solar values of $T_{\rm eff}$$_\odot$ = 5,780 K and $M_{\rm bol}$$_\odot$ = +4.73. For the absolute visual magnitudes ($M_{\rm V}$), 
we used the bolometric corrections (BCs) taken from the expression given by Torres (2010) between $\log T_{\rm eff}$ and BC. 
Using the apparent magnitude of $V$ = +9.36 and the interstellar extinction of $A_{\rm V}$ = 0.097 (Schlafly \& Finkbeiner 2011), 
we obtained the distance to the AI Hya system of 585$\pm$34 pc. This is consistent with 644$\pm$26 pc calculated by the inverted 
parallax from $Gaia$ DR2 (1.554$\pm$0.063 mas; Gaia Collaboration et al. 2018) and 632$\pm$25 pc estimated by the geometric parallax 
(Bailer-Jones et al. 2019) within their errors.

Using the PERIOD04 program, we applied a multifrequency analysis to the eclipse-subtracted light residuals and examined possible 
harmonic and combination frequencies. As a result, 14 frequencies with S/N $>$ 4.0 were found, four ($f_1$, $f_2$, $f_3$, 
and $f_6$) of which may be considered independent pulsations. The pulsation periods for these frequencies are in the range of 
$P_{\rm pul}$ = 0.100$-$0.173 days. The ratios between the pulsation and orbital periods are $P_{\rm pul}/P_{\rm orb}$ = 0.012$-$0.021, 
which is within the upper limit of 0.09 for $\delta$ Sct-type pulsators in binaries (Zhang et al. 2013). Liakos \& Niarchos 
(2017) presented a possible $P_{\rm orb}-P_{\rm pul}$ relation in the binary star systems with orbital periods below 13 days. 
In case of AI Hya, $f_2$ and $f_3$ match well their correlation for the $\delta$ Sct stars in detached EBs, while $f_1$ and $f_6$ do 
not conform to the empirical relation. 
From the well-known relation of $Q$ = $P_{\rm pul}$$\sqrt{\rho / \rho_\odot}$ and the mean density $\rho _2$ given in Table 5, we got 
the observed pulsation constants of $Q_1$ = 0.048 days, $Q_2$ = 0.030 days, $Q_3$ = 0.032 days, and $Q_6$ = 0.052 days, corresponding 
to classical $\delta$ Sct pulsations. Then, we compared the $Q$ values with the theoretical models predicted by Fitch (1981) for 
2.0 $M_\odot$. The $f_2$ and $f_3$ frequencies might be related to the radial fundamental $F$ modes, while the $f_1$ and $f_6$ could 
be regarded as the non-radial $g_1$ modes with a low degree of $\ell$ = 2. AI Hya is a double-lined EB with a multiperiodic 
$\delta$ Sct component, so it should be a good candidate for asteroseismology in binaries. Long-term multiband photometric monitoring 
of the pulsating EB will help to identify the pulsation modes and to inspect the eclipse timing variation in detail.

\begin{ack}
This paper includes data collected by the {\it TESS} mission, which were obtained from MAST. Funding for the {\it TESS} mission 
is provided by the NASA Explorer Program. We thank the {\it TESS} team for its support of this work. This research has made use of 
the Simbad database maintained at CDS, Strasbourg, France, and was supported by the KASI grant 2020-1-830-08. K.H. was supported 
by the grants 2017R1A4A1015178 and 2019R1I1A1A01056776 from the National Research Foundation (NRF) of Korea. 
\end{ack}

\clearpage
\begin{figure}
\includegraphics{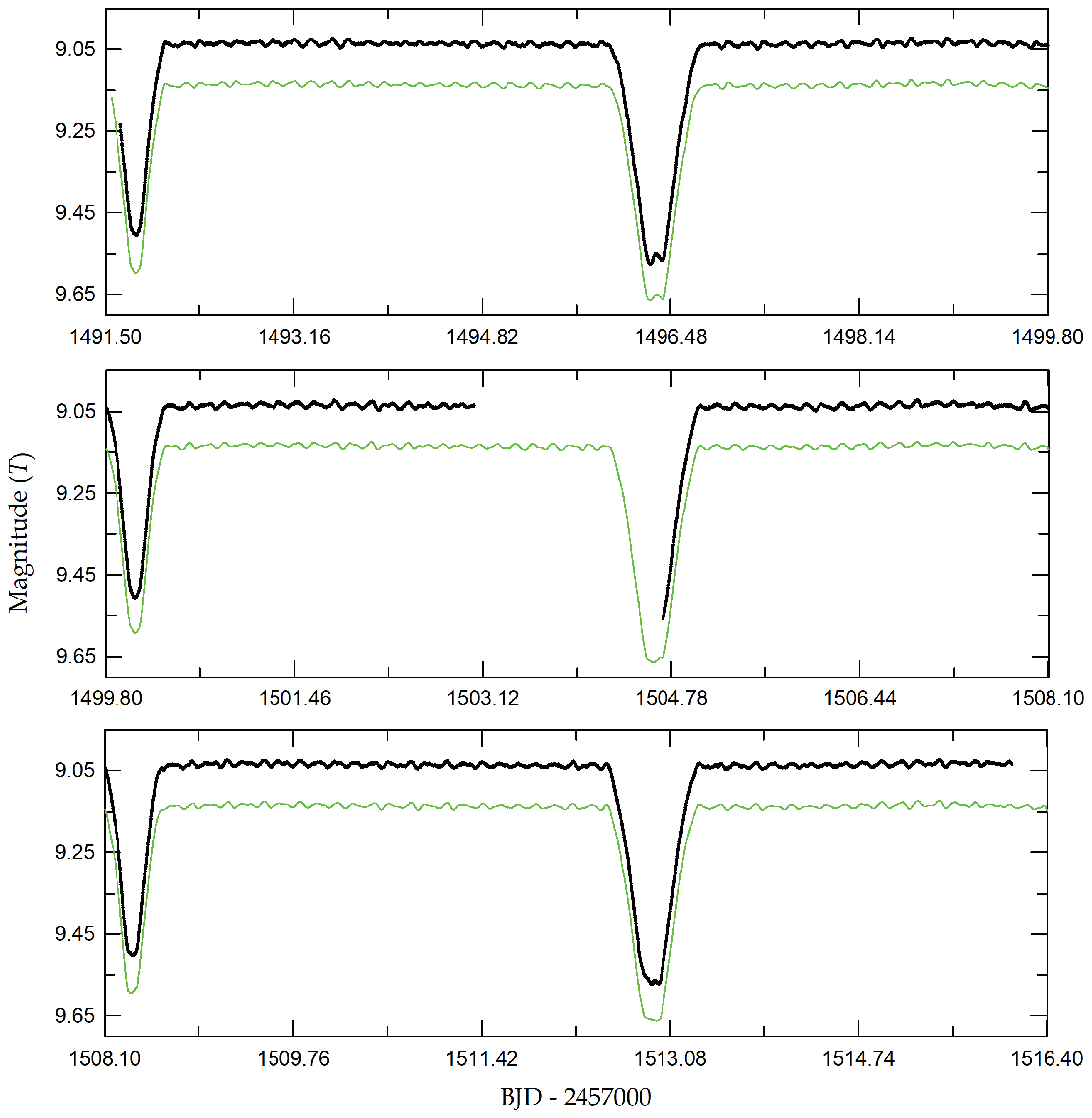}
\caption{Black circles present the {\it TESS} time-series data of AI Hya separated at intervals of 8.3 days. The green lines are 
the sum of the two model curves computed from the Model 2 parameters of Table 3 and the 14-frequency fit of Table 4, respectively. 
They are offset by +0.1 mag for clarity. }
\label{Fig1}
\end{figure}

\begin{figure}
\includegraphics[]{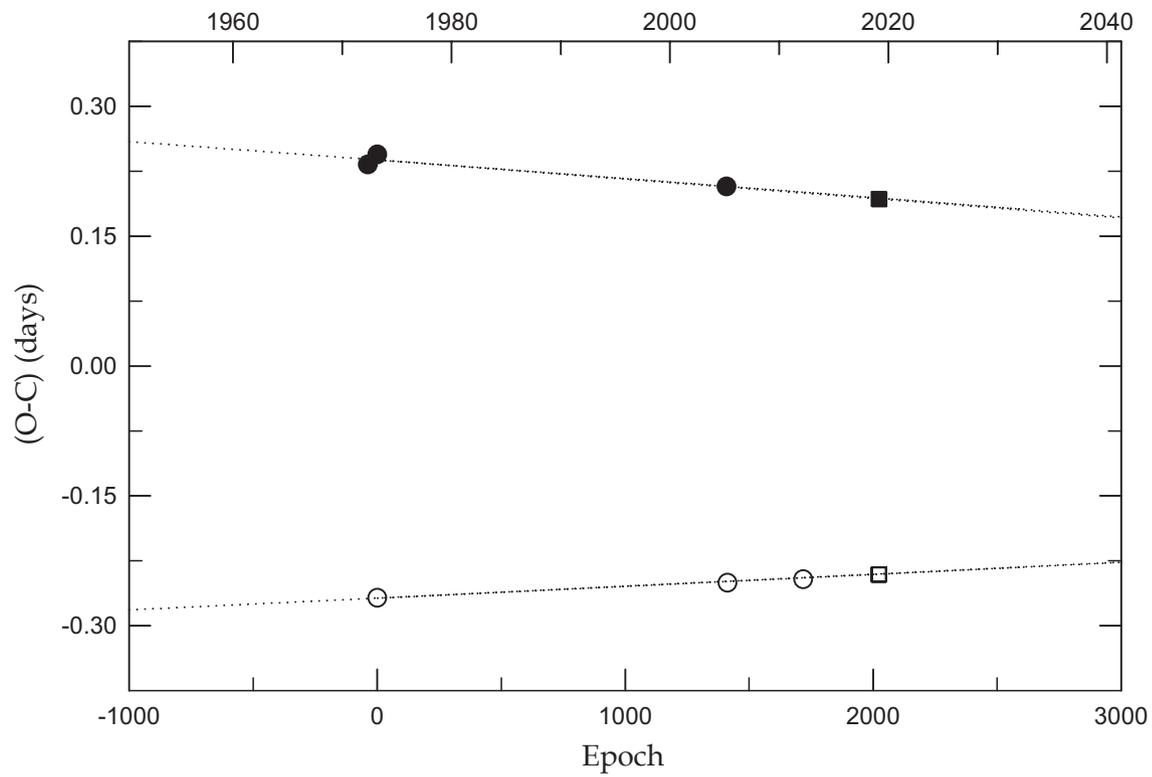}
\caption{Eclipse timing diagram of AI Hya. The filled and open symbols represent the primary and secondary times of minimum light, 
respectively. The circles and squares denote the literature and {\it TESS} eclipse timings, respectively. The dotted lines are computed 
with the apsidal motion elements in Table 2. }
\label{Fig2}
\end{figure}

\begin{figure}
\includegraphics[]{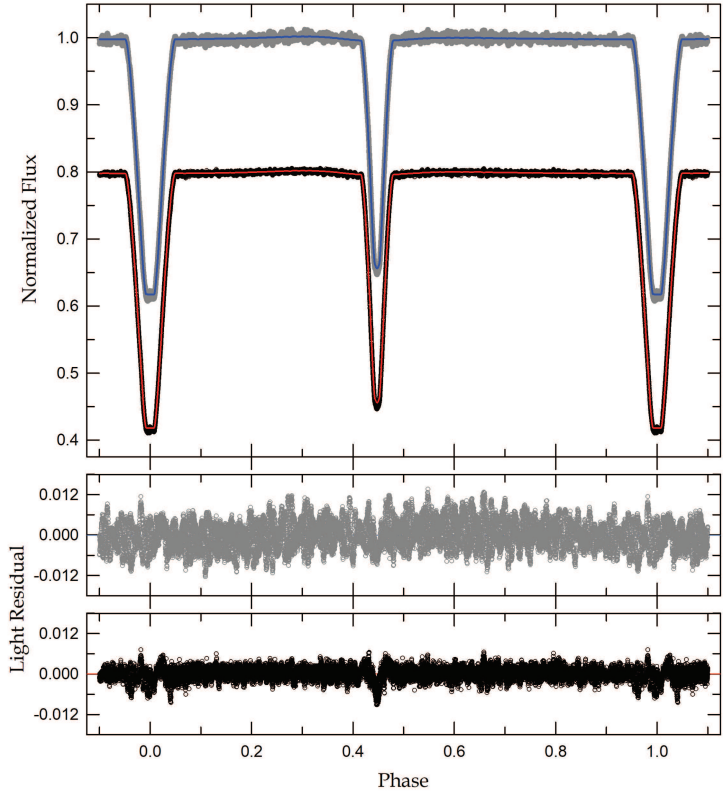}
\caption{Binary light curve before (gray circle) and after (black circle) subtracting the pulsation signatures from the observed 
{\it TESS} data. The blue and red lines are computed with the Model 1 and Model 2 parameters in Table 3, respectively. 
The corresponding residuals from the fits are plotted in the middle and bottom panels in the same order as the light curves. 
In the bottom panel, some features visible during the times of both eclipses may come from insufficient removal of the pulsations 
in the orbital phases, caused by using only out-of-eclipse data in the frequency analysis. }
\label{Fig3}
\end{figure}

\begin{figure}
\includegraphics[scale=0.9]{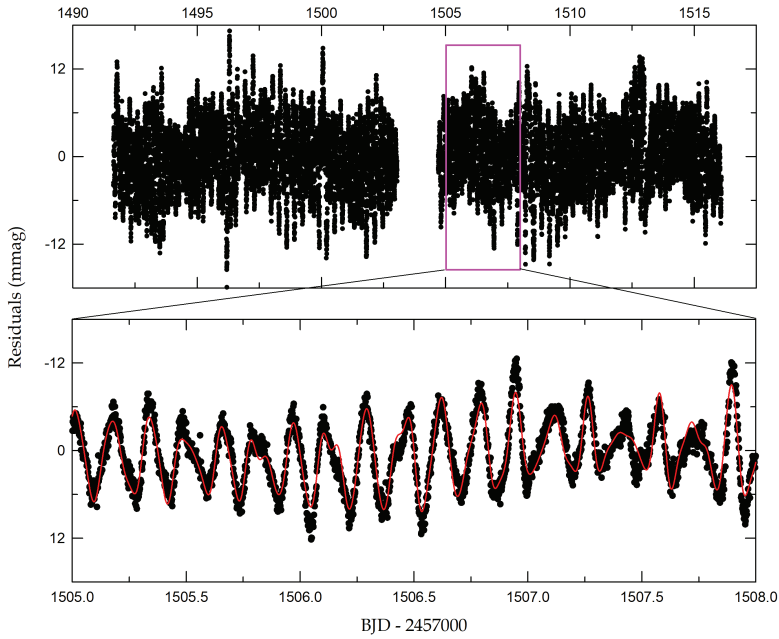}
\caption{Light curve residuals after removing the binarity effects from the observed {\it TESS} data. The lower panel presents a short 
section of the residuals marked by the inset box in the upper panel. The synthetic curve is computed from the 14-frequency fit to 
the whole data except for the times of both eclipses. }
\label{Fig4}
\end{figure}

\begin{figure}
\includegraphics[]{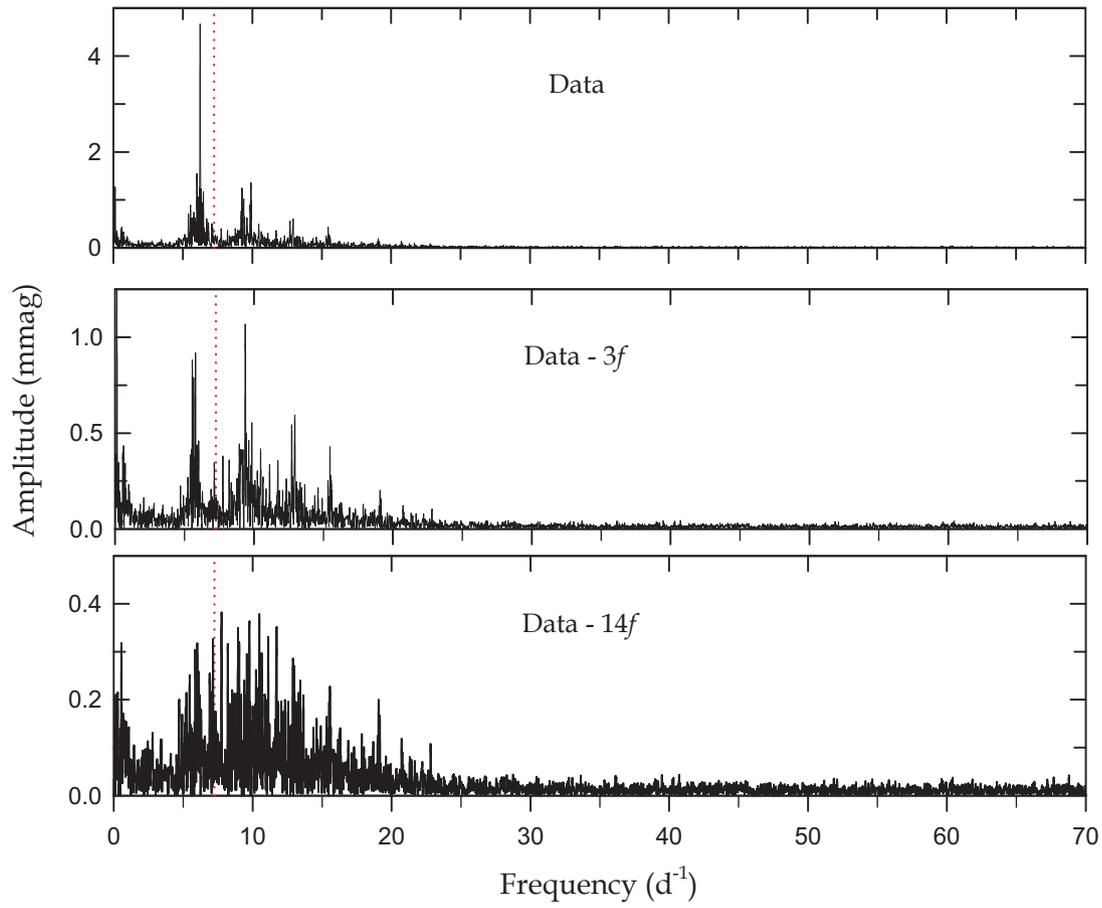}
\caption{Amplitude spectra before (top panel) and after pre-whitening the first three frequencies (middle) and all 14 frequencies 
(bottom) from the PERIOD04 program for the outside-eclipse residuals. In all panels, the red vertical lines represent the location of 
7.2448 day$^{-1}$ previously detected by J\o rgensen \& Gr\o nbech (1978) but undetected in the {\it TESS} data. }
\label{Fig5}
\end{figure}

\clearpage                                                                                                           
\begin{table}
\tbl{Observed photoelectric and CCD times of minimum light for AI Hya. }{%
\begin{tabular}{llrrrcl}
\hline
BJD           & Error         & Epoch           &  $O-C_1$           &  $O-C_2$             &  Min          &  References                         \\ [-2.0ex]
(2,400,000+)  &               &                 &                    &                      &               &                                     \\
\hline
41,411.3685   &               & $-$38.0         &  $+$0.23279        &  $-$0.00623          &  I            &  Kizilirmak \& Pohl (1974)          \\
41,721.7305   &               &  $-$0.5         &  $-$0.26796        &  $+$0.00042          &  II           &  J\o rgensen \& Gr\o nbech (1978)   \\
41,726.3877   &               &     0.0         &  $+$0.24440        &  $+$0.00621          &  I            &  J\o rgensen \& Gr\o nbech (1978)   \\
53,406.5007   & $\pm$0.0007   &  1409.0         &  $+$0.20744        &  $+$0.00043          &  I            &  Kreiner et al. (2001)              \\
53,435.0565   &               &  1412.5         &  $-$0.25062        &  $-$0.00093          &  II           &  Nagai (2006)                       \\
55,963.4110   & $\pm$0.0011   &  1717.5         &  $-$0.24654        &  $-$0.00091          &  II           &  H\"ubscheret al. (2013)            \\
58,491.76677  & $\pm$0.00016  &  2022.5         &  $-$0.24118        &  $+$0.00038          &  II           &  This article (TESS)                \\
58,496.34588  & $\pm$0.00022  &  2023.0         &  $+$0.19309        &  $-$0.00008          &  I            &  This article (TESS)                \\
58,500.05661  & $\pm$0.00005  &  2023.5         &  $-$0.24102        &  $+$0.00053          &  II           &  This article (TESS)                \\
58,508.34627  & $\pm$0.00022  &  2024.5         &  $-$0.24103        &  $+$0.00050          &  II           &  This article (TESS)                \\
58,512.92494  & $\pm$0.00023  &  2025.0         &  $+$0.19280        &  $-$0.00032          &  I            &  This article (TESS)                \\
\hline
\end{tabular}}\label{tab:1}
\end{table}

\begin{table}
\tbl{Apsidal motion elements of AI Hya. }{%
\begin{tabular}{lc}
\hline
Parameter                         &  Value                     \\
\hline
$T_0$ (BJD)                       &  2,441,726.086$\pm$0.017   \\
$P_{s}$ (day)                     &  8.289672$\pm$0.000015     \\
$P_{a}$ (day)                     &  8.289711$\pm$0.000017     \\
$e$                               &  0.241$\pm$0.083           \\
$\omega_{0}$ (deg)                &  247$\pm$8                 \\
$\dot{\omega}$ (deg year$^{-1}$)  &  0.075$\pm$0.031           \\
$U$ (year)                        &  4800$\pm$2000             \\
\hline
\end{tabular}}\label{tab:2}
\end{table}

\begin{table}
\tbl{Binary parameters of AI Hya. }{%
\begin{tabular}{lccccc}
\hline
Parameter                                & \multicolumn{2}{c}{Model 1$\rm ^a$}         && \multicolumn{2}{c}{Model 2$\rm ^b$}        \\ [1.0mm] \cline{2-3} \cline{5-6}
                                         & Primary           & Secondary               && Primary           & Secondary              \\ 
\hline
$T_0$ (BJD)                              & \multicolumn{2}{c}{2,458,496.09280(48)}     && \multicolumn{2}{c}{2,458,496.09319(14)}    \\
$P_{\rm orb}$ (day)                      & \multicolumn{2}{c}{8.29070(31)}             && \multicolumn{2}{c}{8.28994(10)}            \\
$q$                                      & \multicolumn{2}{c}{1.084}                   && \multicolumn{2}{c}{1.084}                  \\
$e$                                      & \multicolumn{2}{c}{0.2317(23)}              && \multicolumn{2}{c}{0.2343(9)}              \\
$\omega$ (deg)                           & \multicolumn{2}{c}{249.73(21)}              && \multicolumn{2}{c}{249.96(9)}              \\
$i$ (deg)                                & \multicolumn{2}{c}{89.314(29)}              && \multicolumn{2}{c}{89.234(18)}             \\
$T_{\rm eff}$ (K)                        & 7309(70)          & 6854(60)                && 7291(33)          & 6864(28)               \\
$\Omega$                                 & 11.280(19)        & 8.759(17)               && 11.241(5)         & 8.816(3)               \\
$\Omega_{\rm in}$                        & \multicolumn{2}{c}{3.885}                   && \multicolumn{2}{c}{3.885}                  \\
$A$                                      & 1.0               & 0.5                     && 1.0               & 0.5                    \\
$g$                                      & 1.0               & 0.32                    && 1.0               & 0.32                   \\
$X$, $Y$                                 & 0.643, 0.249      & 0.635, 0.249            && 0.643, 0.249      & 0.635, 0.249           \\
$x$, $y$                                 & 0.618, 0.268      & 0.620, 0.296            && 0.617, 0.270      & 0.620, 0.296           \\
$L$/($L_{1}$+$L_{2}$)                    & 0.3780(6)         & 0.6220                  && 0.3805(3)         & 0.6195                 \\
$r$ (pole)                               & 0.1012(3)         & 0.1441(23)              && 0.1017(1)         & 0.1431(2)              \\
$r$ (point)                              & 0.1017(3)         & 0.1461(24)              && 0.1022(1)         & 0.1451(2)              \\
$r$ (side)                               & 0.1013(3)         & 0.1445(23)              && 0.1018(1)         & 0.1435(2)              \\
$r$ (back)                               & 0.1017(3)         & 0.1457(23)              && 0.1021(1)         & 0.1447(2)              \\
$r$ (volume)$\rm ^c$                     & 0.1014(3)         & 0.1448(23)              && 0.1019(1)         & 0.1438(2)              \\ 
$\sum W(O-C)^2$                          & \multicolumn{2}{c}{0.0041}                  && \multicolumn{2}{c}{0.0017}                 \\ 
\hline
\end{tabular}}\label{tab:3}
\begin{tabnote}
\footnotemark[a]Result from the observed data.  \\
\footnotemark[b]Result from the pulsation-subtracted data. \\
\footnotemark[c]Mean volume radius.
\end{tabnote}
\end{table}

\begin{table}
\tbl{Results of the multiple frequency analysis for AI Hya.$\rm ^a$ }{%
\begin{tabular}{lrcccc}
\hline
             & Frequency              & Amplitude           & Phase           & S/N$\rm ^b$    & Remark$\rm ^c$                  \\ [-2.0ex]
             & (day$^{-1}$)           & (mmag)              & (rad)           &                &                                 \\
\hline
$f_{1}$      &  6.2406$\pm$0.0002     & 4.76$\pm$0.16       & 2.58$\pm$0.10   & 51.41          &                                 \\
$f_{2}$      &  9.9064$\pm$0.0012     & 1.21$\pm$0.19       & 1.26$\pm$0.47   & 10.76          &                                 \\
$f_{3}$      &  9.2539$\pm$0.0017     & 0.93$\pm$0.20       & 2.71$\pm$0.62   &  8.07          &                                 \\
$f_{4}$      &  0.1205$\pm$0.0007     & 1.37$\pm$0.12       & 4.25$\pm$0.26   & 18.98          & $f_{\rm orb}$                   \\
$f_{5}$      &  9.3682$\pm$0.0012     & 1.31$\pm$0.20       & 5.20$\pm$0.44   & 11.30          & $f_3+f_{\rm orb}$               \\
$f_{6}$      &  5.7834$\pm$0.0017     & 0.69$\pm$0.15       & 4.92$\pm$0.63   &  7.91          &                                 \\
$f_{7}$      &  5.5548$\pm$0.0015     & 0.77$\pm$0.14       & 3.92$\pm$0.55   &  9.07          & $f_1+f_3-f_2$                   \\
$f_{8}$      &  5.6379$\pm$0.0016     & 0.70$\pm$0.15       & 0.82$\pm$0.61   &  8.25          & $f_6-f_{\rm orb}$               \\
$f_{9}$      &  9.8441$\pm$0.0023     & 0.66$\pm$0.19       & 1.30$\pm$0.85   &  5.88          & $f_2-f_{\rm orb}$               \\
$f_{10}$     & 12.9322$\pm$0.0024     & 0.55$\pm$0.16       & 3.36$\pm$0.88   &  5.71          & $f_2+f_3-f_1$                   \\
$f_{11}$     &  9.3183$\pm$0.0022     & 0.73$\pm$0.20       & 3.09$\pm$0.79   &  6.32          & $f_5$                           \\
$f_{12}$     &  9.2040$\pm$0.0029     & 0.54$\pm$0.19       & 3.37$\pm$1.05   &  4.79          & $f_3$                           \\
$f_{13}$     & 12.7161$\pm$0.0028     & 0.48$\pm$0.17       & 4.54$\pm$1.02   &  4.89          & $2f_1+2f_{\rm orb}$             \\
$f_{14}$     & 15.4634$\pm$0.0023     & 0.42$\pm$0.12       & 3.99$\pm$0.84   &  5.98          & $f_1+f_3$                       \\
\hline
\end{tabular}}\label{tab:4}
\begin{tabnote}
\footnotemark[a]Uncertainties were calculated according to Kallinger et al. (2008). \\
\footnotemark[b]Calculated in a range of 5 day$^{-1}$ around each frequency. \\
\footnotemark[c]Possible harmonic and combination frequencies. \\
\end{tabnote}
\end{table}

\begin{table}
\tbl{Absolute parameters of AI Hya. }{%
\begin{tabular}{lccccc}
\hline
Parameter                   & \multicolumn{2}{c}{Popper (1988)}        && \multicolumn{2}{c}{This Paper}          \\ [1.0mm] \cline{2-3} \cline{5-6} 
                            & Primary           & Secondary            &&  Primary           & Secondary          \\
\hline
$M$ ($M_\odot$)             & 1.98$\pm$0.04    & 2.15$\pm$0.04         && 1.97$\pm$0.03    & 2.14$\pm$0.03        \\
$R$ ($R_\odot$)             & 2.77$\pm$0.02    & 3.92$\pm$0.03         && 2.81$\pm$0.02    & 3.96$\pm$0.04        \\
$\log$ $g$ (cgs)            & 3.85$\pm$0.01    & 3.58$\pm$0.01         && 3.84$\pm$0.01    & 3.57$\pm$0.01        \\
$\rho$ ($\rho_\odot$)       &                  &                       && 0.089$\pm$0.002  & 0.034$\pm$0.001      \\
$T_{\rm eff}$ (K)           & 7096$\pm$65      & 6699$\pm$62           && 7291$\pm$200     & 6864$\pm$200         \\
$L$ ($L_\odot$)             & 17.4$\pm$0.8     & 27.5$\pm$1.3          && 20$\pm$2         & 31$\pm$4             \\
$M_{\rm bol}$ (mag)         &                  &                       && 1.48$\pm$0.12    & 0.99$\pm$0.13        \\
BC (mag)                    &                  &                       && 0.03$\pm$0.01    & 0.03$\pm$0.01        \\
$M_{\rm V}$ (mag)           & 1.53$\pm$0.04    & 1.05$\pm$0.04         && 1.45$\pm$0.12    & 0.97$\pm$0.13        \\
Distance (pc)               & \multicolumn{2}{c}{575$\pm$15}           && \multicolumn{2}{c}{585$\pm$34}          \\
\hline
\end{tabular}}\label{tab:5}
\end{table}

\end{document}